\documentclass[a4paper]{article}
\usepackage{multirow}
\usepackage{booktabs}
\usepackage{INTERSPEECH2020}
\usepackage{stfloats}
\usepackage{color}
\newcommand{\tabincell}[2]{\begin{tabular}{@{}#1@{}}#2\end{tabular}}

\title{The INTERSPEECH 2020 Far-Field Speaker Verification Challenge}
\name{Xiaoyi Qin$^{1}$, Ming Li$^{1,5}$, Hui Bu$^{4}$, Wei Rao$^{2}$, Rohan Kumar Das$^{2}$,\\ Shrikanth Narayanan$^{3}$, Haizhou Li$^{2}$}

\address{$^{1}$Data Science Research Center, Duke Kunshan University, Kunshan, China \\
            $^{2}$Department of Electrical \& Computer Engineering, National University of Singapore, Singapore \\
            $^{3}$Signal Analysis and Interpretation Lab, University of Southern California, Los Angeles, USA \\
             $^{4}$AI Shell Foundation, Beijing, China \\
             $^{5}$School of Computer Science, Wuhan University, Wuhan, China} 
             
\email{ming.li369@dukekunshan.edu.cn}

\begin{document}

\maketitle

\begin{abstract}
The INTERSPEECH 2020 Far-Field Speaker Verification Challenge (FFSVC 2020) addresses three different research problems under well-defined conditions: far-field text-dependent speaker verification from single microphone array, far-field text-independent speaker verification from single microphone array, and far-field text-dependent speaker verification from distributed microphone arrays. All three tasks pose a cross-channel challenge to the participants. To simulate the real-life scenario, the enrollment utterances are recorded from close-talk cellphone, while the test utterances are recorded from the far-field microphone arrays. In this paper, we describe the database, the challenge, and the baseline system, which is based on a ResNet-based deep speaker network with cosine similarity scoring. For a given utterance, the speaker embeddings of different channels are equally averaged as the final embedding. The baseline system achieves minDCFs of 0.62, 0.66, and 0.64 and EERs of 6.27\%, 6.55\%, and 7.18\% for task 1, task 2, and task 3, respectively.
\end{abstract}


\section{Introduction}
Automatic speaker verification (ASV) is an enabling technology in speech processing and biometric authentication. It has been deployed in many real-life applications, such as access control, and law enforcement. With the advent of deep learning, speaker recognition performance has improved remarkably in both close-talk and far-field settings. However, it is still far from perfect, for example, speaker recognition under noisy and far-field conditions remains a challenging task. The INTERSPEECH 2020 Far-Field Speaker Verification Challenge aims at providing a common platform for the research community to advance the state-of-the-art.

A typical deep speaker network firstly learns frame-level speaker representation with the local pattern extractor, which is usually a time-delayed neural network (TDNN) \cite{xvector_icassp} or a convolutional neural network (CNN)~\cite{cai_lde}. The learnt frame-level feature sequence is then converted into a fixed-dimension representation by different pooling mechanisms such as statistics pooling \cite{xvector_icassp}, attentive pooling \cite{selfattention_speaker}, and learnable dictionary encoding \cite{cai_lde}. Since speaker verification in the open set settings is essentially a metric learning problem, several discriminative classification losses such as A-softmax \cite{a_softmax} and AM-softmax \cite{am_softmax} are employed to enhance the recognition performance.
 
To compensate for the adverse impacts of reverberation and noise in the far-field scenario, various approaches have been proposed for ASV systems. At signal level, weighted prediction error \cite{wpe_1,wpe_2} is employed for dereverberation. DNN-based denoising~\cite{sednn, segan, SE_lstm} and beamforming \cite{nn_gev_experment, nn_gev} are investigated for single-channel and multi-channel speech enhancement respectively. At the modeling level, data augmentation \cite{Cai2019, STC_voices19, BUT_voices19} and transfer learning~\cite{xiaoyi_farfield} are proven to be effective with limited target domain data. To learn a noise-invariant speaker embedding, adversarial training \cite{zhou_training_2019, meng_adversarial_2019} and variability-invariant loss~\cite{danwei_icassp20_loss} are investigated. Also, joint training of speech enhancement network and speaker embedding network can improve the ASV performance under noisy conditions \cite{jl_se_sv, shon_voiceid_2019, zhao_robust_2019}. For deep speaker modeling with microphone array, a multi-channel training framework is proposed for speaker embedding extraction~\cite{danwei_3d_asv}. Moreover, in the testing phase, enrollment data augmentation is proposed to reduce the mismatch between the enrollment and testing utterances \cite{xiaoyi_farfield}.

Recently, far-field speaker recognition attracts more and more attention from the research community. The Voices Obscured in Complex Environmental Settings (VOiCES) Challenge launched in 2019 aims to benchmark state-of-the-art speech processing methods in far-field and noisy conditions~\cite{VOiCES}. The wake-up word dataset \textit{Hi Mia} has also been released to facilitate the studies in far-field speaker recognition~\cite{himia_dataset}. However, some research questions still require further exploration for speaker verification in the far-field and complex environments. Those open challenges including but not limited to,

\begin{enumerate}
    \item{Far-field text-dependent speaker verification for wake up control}
    \item{Far-field text-independent speaker verification with complex environments}
    \item{Far-field speaker verification with cross-channel enrollment and test}
    \item{Far-field speaker verification with single multi-channel microphone array}
    \item{Far-field speaker verification with multiple distributed microphone arrays}
    \item{Far-field speaker verification with front-end speech enhancement methods}
    \item{Far-field speaker verification with end-to-end modeling using data augmentation}
    \item{Far-field speaker verification with front-end and back-end joint modeling}
    \item{Far-field speaker verification with transfer learning and domain adaptation}
\end{enumerate}
 
To this end, we collect a large scale far-field speaker verification dataset with real speakers in multiple scenarios, which include text-dependent, text-independent, cross channel enrollment and test, distributed microphone array, etc. We also launch the Far-Field Speaker Verification Challenge 2020 (FFSVC 2020) based on this dataset. It focuses on far-field distributed microphone arrays under noisy conditions in real scenarios. The objectives of this challenge are to: 1) benchmark the current speech verification technology under this challenging condition, 2) promote the development of new ideas and techniques in speaker verification, 3) provide an open, free, and large scale speech dataset to the community that exhibits the far-field characteristics in real scenes. 

The challenge consists of three tasks with  different setups,
\begin{itemize}
\item{Task 1: far-field text-dependent speaker verification from single microphone array}
\item{Task 2: far-field text-independent speaker verification from single microphone array}
\item{Task 3: far-field text-dependent speaker verification from distributed microphone arrays}
\end{itemize}

All three tasks pose a cross-channel challenge, that is to have enrollment speech from the close-talk cellphone, and to have test speech from far-field microphone array(s). This paper provides the description of the challenge, the dataset, and the reference baseline.  


\section{Challenge Dataset}

\subsection{DMASH Dataset}

\begin{figure}[t]
  \centering
  \includegraphics[width=0.9\linewidth]{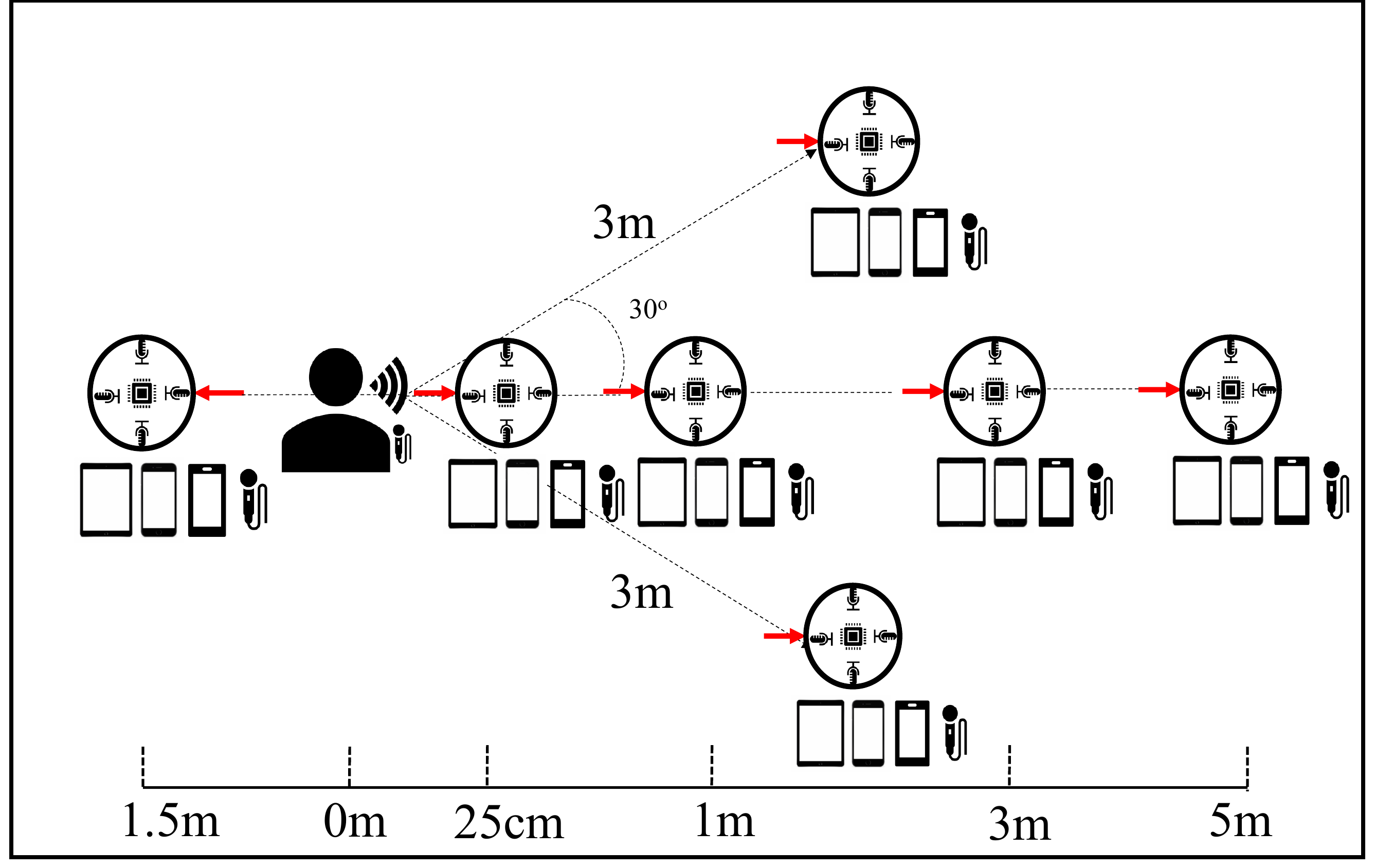}
  \caption{The  setup of the recording environment}
  \label{fig:f1}
\end{figure}

The Distributed Microphone Arrays in Smart Home (DMASH) dataset is recorded in real smart home scenarios with two different rooms. Figure \ref{fig:f1} shows the recording setup of DMASH dataset. The recording devices include one close-talk microphone and seven groups of devices at seven different positions of the room. A group of recording devices include one iPhone, one Android phone, one iPad, one microphone, and one circular microphone array with a radius of 5cm. The red arrow in figure 1 points to channel 0 of microphone arrays. 

During data collection, each speaker visits three times with a gap of 7-15 days. In the first visit (F), the noise sources include an electric fan and the TV broadcast or the office ambient noise. The recording environment of the second visit (S) is quiet. In the third visit (T), the electric fan is the only noise. In each visit, more than 300 utterances for each speaker are recorded. The first 30 utterances are of fixed content: {\textit{`ni hao mi ya'}} in Mandarin Chinese. The next 60 utterances are defined as semi-text-dependent, in which the text content is {\textit{`ni hao mi ya'}} followed by some random text. The remaining utterances are text-independent. All speech recordings are in Mandarin Chinese.

\subsection{FFSVC 2020 Challenge Dataset}

The FFSVC 2020 challenge dataset is part of the DMASH dataset. It includes the recordings from the close-talk microphone, the iPhone at 25cm distance, and three randomly selected circular microphone arrays. For the circular microphone arrays, only four recording channels are used. Under this data selection protocol, each utterance have 14 (1 + 1 + 4 $\times$ 3) recording channels.

In FFSVC 2020, the training partition includes 120 speakers and the development partition includes 35 speakers. Additionally, any publicly open and freely accessible dataset shared on \texttt{openslr} before Feb. $1^{\text{st}}$, 2020 can be used for training\footnote{Dataset published on \texttt{openslr} before SLR85, including SLR85.}.

Table \ref{tab:data details} shows the details of the challenge data. More information about the dataset can be found in \cite{eval_plan}.

\begin{table}[t]
  \caption{The summary of the FFSVC20 challenge data}
  \label{tab:data details}
  \centering
  \begin{tabular}[c]{ll|l}
    \toprule
  Utt ID&  Content & Noise \\
    \midrule
  \multirow{2}*{001-030}&  \textit{ni hao mi ya} & F: TV/Office + electric fan  \\
  &(text-dependent) & T: electric fan \\
  091-  & text independent & S: quiet \\
  \bottomrule
  \end{tabular}
\end{table}

For each task, the evaluation data includes 80 speakers. There is no overlapping among the speakers in the training, development, and evaluation sets. Moreover, there is no overlapping among the evaluation data of the three tasks. Recordings from the iPhone at 25cm distance are selected for enrollment. For testing, one microphone array is used in task 1 and task 2; 2-4 microphone arrays are randomly selected in task 3. For each true trial, the enrollment and the testing utterance are from different visits of the same speaker.

\section{The Baseline System}
\subsection{Data Processing}
\subsubsection{Data augmentation} \label{sec:da}
To improve the robustness and generalization of the deep speaker network, we used \texttt{pyroomacoustics} toolkit \cite{pyroomacoustics} to simulate the room acoustic and generate far-field training data. The room width was randomly set between six to eight meters, and the locations of the speaker, noise, and microphones were also randomly distributed. The noise sources were from MUSAN dataset \cite{musan}, and the signal-to-noise-ratio (SNR) was between 0 to 20 dB. 

\subsubsection{Voice activity detection}
Two voice activity detection (VAD) methods were explored in the systems. The first one was the energy-based VAD. The second method was the gradient boosting algorithm-based voice activity detection (GVAD) \cite{catboost} for the far-field speeches. GVAD is a classifier that separates the speech segments from the non-speech segments. The training data of GVAD was the simulated far-field speeches from the AISHELL-1 dataset (SLR33) \cite{aishell_1}, as described in Section \ref{sec:da}. The `speech' and `non-speech' labels are generated with an energy-based VAD on the original clean data of SLR33. All the far-field speeches of FFSVC 2020 dataset are processed with the trained GVAD before testing.
 
\subsection{Acoustic Feature Extraction}
Audios were resampled to 16,000 Hz and pre-emphasized before feature extraction. Two acoustic features were used: (1) 64-dimensional log Mel-filterbank energies with a frame length of 25ms and hop size of 10ms and (2) 30-dim MFCCs. The former features were used for ResNet-34 and the latter features were used for ResNet-50. The extracted features were mean-normalized before feeding into the deep speaker network.

\begin{table}[t]
    \footnotesize
    \caption{ The ResNet-34 network architecture, $\mathbf{C}$(kernal size, stride) denotes the convolutional layer, $\mathbf{S}$(kernal size, stride) denotes the shortcut convolutional layer, $\left[\cdot \right]$ denotes the residual block.}
    \centering
    \begin{tabular}[c]{@{\ \ \ }l@{\ \ \ }c@{\ \ \ }l@{\ \ \ }}
        \toprule
        \textbf{Layer} & \textbf{Output Size} & \textbf{Structure} \\
        \midrule
        Conv1 & $32 \times 64 \times L$ & $\mathbf{C}(3\times 3, 1)$ \\
        \midrule
        \tabincell{l}{Residual\\Layer 1} & $32 \times 64 \times L$ & $\begin{bmatrix}
            \mathbf{C}(3\times 3, 1) \\
            \mathbf{C}(3\times 3, 1)
        \end{bmatrix} \times 3$ \\
        \midrule
        \tabincell{l}{Residual\\Layer 2} & $64 \times 32 \times \frac{L}{2}$ & $\begin{bmatrix}
            \mathbf{C}(3\times 3, 2) \\
            \mathbf{C}(3\times 3, 1) \\
            \mathbf{S}(1\times 1, 2)
        \end{bmatrix} \begin{bmatrix}
            \mathbf{C}(3\times 3, 1) \\
            \mathbf{C}(3\times 3, 1)
        \end{bmatrix}\times 3$ \\
        \midrule
        \tabincell{l}{Residual\\Layer 3} & $128 \times 16 \times \frac{L}{4}$ & $\begin{bmatrix}
            \mathbf{C}(3\times 3, 2) \\
            \mathbf{C}(3\times 3, 1) \\
            \mathbf{S}(1\times 1, 2)
        \end{bmatrix} \begin{bmatrix}
            \mathbf{C}(3\times 3, 1) \\
            \mathbf{C}(3\times 3, 1)
        \end{bmatrix}\times 5$ \\
        \midrule
        \tabincell{l}{Residual\\Layer 4} & $256 \times 8 \times \frac{L}{8}$ & $\begin{bmatrix}
            \mathbf{C}(3\times 3, 2) \\
            \mathbf{C}(3\times 3, 1) \\
            \mathbf{S}(1\times 1, 2)
        \end{bmatrix} \begin{bmatrix}
            \mathbf{C}(3\times 3, 1) \\
            \mathbf{C}(3\times 3, 1)
        \end{bmatrix}\times 2$ \\
        \midrule
        Encoding & $512$ & Global Statistics Pooling \\
        \midrule
        Embedding & $128$ & Fully Connected Layer\\
        Classifier & $10544$ & Fully Connected Layer\\
        \bottomrule
    \end{tabular}
    \label{table: architecture}
\end{table}

\begin{table}[t]
	\label{table: architecture_resnet50}
    \footnotesize
    \caption{The ResNet-50 network architecture. The meaning of parameters in this table can be referred to the caption of Table \ref{table: architecture}.}
    \centering
    \begin{tabular}[c]{@{\ \ \ }l@{\ \ \ }c@{\ \ \ }l@{\ \ \ }}
        \toprule
        \textbf{Layer} & \textbf{Output Size} & \textbf{Structure} \\
        \midrule
        Conv1 & $64 \times 128 \times L$ & $\mathbf{C}(7\times 7, 2)$ \\
        \midrule
        \tabincell{l}{Residual\\Layer 1} & $256 \times 64 \times \frac{L}{2}$ & $\begin{bmatrix}
            \mathbf{C}(1\times 1, 1) \\
            \mathbf{C}(3\times 3, 2) \\
            \mathbf{S}(1\times 1, 1)
        \end{bmatrix} \times 3$ \\
        \midrule
        \tabincell{l}{Residual\\Layer 2} & $512 \times 32 \times \frac{L}{4}$ & $\begin{bmatrix}
            \mathbf{C}(1\times 1, 1) \\
            \mathbf{C}(3\times 3, 2) \\
            \mathbf{S}(1\times 1, 1)
        \end{bmatrix} \times 4$ \\
        \midrule
        \tabincell{l}{Residual\\Layer 3} & $1024 \times 16 \times \frac{L}{8}$ & $\begin{bmatrix}
            \mathbf{C}(1\times 1, 1) \\
            \mathbf{C}(3\times 3, 2) \\
            \mathbf{S}(1\times 1, 1)
        \end{bmatrix}\times 6$ \\
        \midrule
        \tabincell{l}{Residual\\Layer 4} & $2048 \times 8 \times \frac{L}{16}$ & $\begin{bmatrix}
            \mathbf{C}(1\times 1, 1) \\
            \mathbf{C}(3\times 3, 2) \\
            \mathbf{S}(1\times 1, 1)
        \end{bmatrix}\times 5$ \\
        \midrule
        Encoding & $2048$ & Global Statistics Pooling \\
        \midrule
        Embedding & $1024$ & Fully Connected Layer\\
        Classifier & $2447$ & Fully Connected Layer\\
        \bottomrule
    \end{tabular}
\end{table}

\subsection{Deep Speaker Embedding}

ResNet based networks were applied for FFSVC. Two different ResNet architectures were investigated: (1) ResNet-34 and (2) ResNet-50.

\subsubsection{ResNet-34} \label{sec:res34}
The network structure contains three main components: a front-end pattern extractor, an encoding layer, and a back-end classifier. The ResNet-34 structure \cite{resnet_he} is adopted as the front-end pattern extractor. It learns a frame-level representation from the input spectral features. The global statistics pooling (GSP) layer is then used as the encoder layer to compute the mean and standard deviation of the input frame-level feature sequence. The GSP layer outputs an utterance-level representation with speaker information. A fully-connected layer with a classification output layer then processes the utterance-level representation. Each unit in the output layer is represented as a target speaker identity. All the components in the pipeline are jointly learned with cross-entropy loss. The detailed configuration of the neural network is in Table \ref{table: architecture}.

We pre-trained the deep speaker network with larger scale text-independent mix-dataset (close-talk and its simulation data). The pre-training data contained 10,554 speakers, including SLR33, SLR38, SLR47, SLR49, SLR62, and SLR68 from \texttt{openslr.org}. In the pre-training stage, the model was trained for 50 epochs with an initial learning rate of 0.1. The learning rate was divided by ten every 20 epochs. The network was optimized by stochastic gradient descent.

According to previous works, fine-tuning was an effective transfer learning approach for far-field ASV \cite{xiaoyi_farfield}. In task 1 and task 3, the fine-tuning data is SLR85 dataset and the first 30 utterances of FFSVC 2020 training dataset. The remaining FFSVC 2020 training dataset is used to fine-tune the model for task 2. To prevent overfitting during fine-tuning, data augmentation is also employed to simulate the far-field data for the clean close-talk channel. The real and simulated far-field data jointly fine-tune the pre-trained model. The learning rate is set to 0.001 when fine-tuning. 

\subsubsection{ResNet-50}
The network structure of ResNet-50 is similar as ResNet-34 and also composed of three main components described in Section \ref{sec:res34}. The details of configuration is shown in Table \ref{table: architecture_resnet50}.

SLR33, SLR38, SLR62, SLR82, SLR85 from \texttt{openslr.org} and FFSVC training set were used for training. The training data contained 2,447 speakers. The model was trained for 25 epochs with an initial learning rate of 0.1. The learning rate was divided by ten every 5 epochs. The network was optimized by stochastic gradient descent.

\begin{table*}[t]
  \caption{Performance of the speaker verification systems. ``Model'' represents the types of deep speaker embedding network; ``Scoring'' represents the back-end scoring; ``DA'' represents whether performing data augmentation; ``Enrollment'' represents whether some methods were applied for enrollment utterances; ``Test'' represents the types of utterances selected as test for scoring; ``Single'' represents that utterance from one channel of microphone array(s) is selected as test for scoring; ``Multi'' represents that the utterances of all channels from the microphone array(s) are used as test for scoring.}
  \label{tab:t1}
  \centering
  \resizebox{\textwidth}{15mm}{
  \begin{tabular}[c]{@{\ }l@{\ \ }l@{\ \ }l@{\ \ }l@{\ \ }l@{\ \ }l@{\ \ }c@{\ \ }c@{\ \ }c@{\ \ }c@{\ \ }c@{\ \ }c@{\ \ }c@{\ \ \ }c@{\ \ }c@{\ \ }c@{\ \ }c@{\ \ }c@{\ \ }c@{\ }}
    \toprule
     & & & & &  \multicolumn{6}{c}{\textbf{Development Set}} & \multicolumn{6}{c}{\textbf{Evaluation Set}} \\
     \cmidrule(lr){7-12} \cmidrule(lr){13-18} 
      \multirow{3}*{\textbf{ID}} & \multirow{3}*{\textbf{Model}} &
      \multirow{3}*{\textbf{Scoring}} &
      \multirow{3}*{\textbf{DA}} &
      \multirow{3}*{\textbf{Enrollment}} & \multirow{3}*{\textbf{Test}} &  \multicolumn{2}{c}{\textbf{Task1}} & \multicolumn{2}{c}{\textbf{Task2}} & \multicolumn{2}{c}{\textbf{Task3}} & \multicolumn{2}{c}{\textbf{Task1}} & \multicolumn{2}{c}{\textbf{Task2}} & \multicolumn{2}{c}{\textbf{Task3}} \\
      \cmidrule(lr){7-8} \cmidrule(lr){9-10} \cmidrule(lr){11-12} \cmidrule(lr){13-14} \cmidrule(lr){15-16} \cmidrule(lr){17-18} 
      & & & & & & minDCF & EER & minDCF & EER & minDCF & EER & minDCF & EER & minDCF & EER & minDCF & EER  \\

    \midrule
     $1$ & ResNet50 & CDS & N & -- & Multi & 0.86 & 8.04\%  & 0.97 & 10.96\%  & 0.88 & 7.22\% & 0.87 & 9.93\% & 0.95 & 11.87\%  & 0.86 & 10.68\% \\     
     $2$ & ResNet34 & CDS & Y & -- & Single & 0.64 & 6.30\%  & 0.65 & 6.23\%  & 0.64 & 5.82\% &  0.71  & 7.02\%  & 0.72 & 6.93\%  & 0.68 & 7.78\% \\
     \multirow{2}*{$3$} & ResNet34 & CDS & Y & \multirow{2}*{--} & \multirow{2}*{Multi} & \multirow{2}*{\bf{0.57}}  & \multirow{2}*{6.01\%} & \multirow{2}*{\bf{0.58}} & \multirow{2}*{5.83\%} & \multirow{2}*{\bf{0.59}} & \multirow{2}*{5.42\%} & \multirow{2}*{\bf{0.62}}  & \multirow{2}*{6.37\%} & \multirow{2}*{\bf{0.66}} & \multirow{2}*{6.55\%} & \multirow{2}*{\bf{0.64}} & \multirow{2}*{7.18\%} \\
     ~ & \bf{(Baseline} & \bf{System)} \\
     $4$ & ResNet34 & PLDA & Y & -- & Multi & 0.58 & 5.92\% & 0.60 & 5.69\% & 0.61 & 5.36\% & 0.63  & 6.28\% & 0.67 & 6.48\% & 0.67 & 7.10\% \\
     $5$ & ResNet34 & CDS & Y & EDA & Multi & 0.60 & 5.87\% & 0.61 & 5.61\% & 0.60 & 5.33\% &  0.64  & 6.23\% & 0.68 & 6.36\% & 0.71 & 7.03\% \\   
     \bottomrule 
     \end{tabular}}
\end{table*}

\subsection{Back-end Scoring}
The cosine distance scoring (CDS) and probabilistic linear discriminant analysis (PLDA) \cite{Kenny10, Prince07} served as the back-end scoring methods.

\subsection{Enrollment Data Augmentation}

In far-field speaker verification, the mismatch between enrollment and testing utterances generally exists due to the different recording environments. Data augmentation on enrollment utterances is proven to be effective in reducing this mismatch \cite{himia_dataset}. In this paper, instead of randomly simulating the far-field enrollment data, we used the background noise of the testing utterance to perform enrollment augmentation. Specifically, a GVAD was adopted to detect the non-speech parts of the testing utterance for each trial. These non-speech parts were used as the background noise to get a simulated enrollment utterance. The speaker embeddings from the simulated and original utterance were equally weighted to get the final enrollment embedding.

\section{Experiment Results}

Table \ref{tab:t1} shows the performance of the speaker verification systems under development and evaluation condition, respectively. The performance metrics for FFSVC 2020 are equal error rate (EER) and minimum detection cost function (minDCF) with $P_{\textrm{target}}=0.01$. The minDCF is used as primary metric.

Experimental results of System ID 1 and ID 3 in Table \ref{tab:t1} show that larger scale of training data and data augmentation could significantly improve the performance while ResNet34 is shallower than ResNet50.

During testing, different channels from the microphone array(s) are equally weighted at the embedding level before scoring(ID 3 in table \ref{tab:t1}). One channel of the microphone array(s) is selected as single-channel testing for comparison (ID 2 in table \ref{tab:t1}). The results of enrollment data augmentation (EDA) are also given. Additional attention should be paid to our results in metric: despite the results of PLDA and EDA are better in terms of EER, the minDCF is not so that. Finally, the embedding level averaging model(ID 3) which achieves the best single-system results on evaluation and development datasets is selected as baseline system for FFSVC 2020 challenge.

\section{Conclusions}

The primary purpose of the FFSVC 2020 is to investigate how well the speaker verification technology processes the real-world audio data, especially for the far-field distributed microphone arrays. The challenge data will be released as a large scale speech database after the competition. This paper also provides the description of the baseline system. We believe that this challenge and the published corpus will promote the advancement of research and technology development in far-field speaker recognition.

\section{Acknowledgements}

This research is funded in part by the National Natural Science Foundation of China (61773413), Key Research and Development Program of Jiangsu Province (BE2019054), Six talent peaks project in Jiangsu Province (JY-074), Science and Technology Program of Guangzhou, China (202007030011, 201903010040), and also supported by Programmatic Grant No. A1687b0033 from the Singapore Government's Research, Innovation and Enterprise 2020 plan (Advanced Manufacturing and Engineering domain), and Human-Robot Interaction Phase 1 (Grant No. 192 25 00054) by the National Research Foundation, Prime Minister's Office, Singapore under the National Robotics Programmer.

\bibliographystyle{IEEEtran}

\bibliography{template}

\end{document}